\documentclass[amsmath, amssymb, preprintnumbers, showpacs, showkeys, aps, prl, twocolumn]{revtex4-1}

\usepackage{graphicx}
\usepackage{float}
\usepackage{color}
\usepackage{float}
\usepackage{color} 
\usepackage{wrapfig}

\usepackage{mathtools} 
\usepackage{extarrows} 
\usepackage[titles,subfigure]{tocloft} 

\usepackage{graphicx} 

\usepackage{booktabs} 
\usepackage{array} 
\usepackage{paralist} 
\usepackage{verbatim} 
\usepackage{subfig} 
\usepackage{latexsym}
\usepackage{esint}
\usepackage{geometry} 
\usepackage[utf8]{inputenc} 

\usepackage{mathrsfs}
\usepackage{amsmath, amsthm, amssymb}
\usepackage{color}
\usepackage{epstopdf}

\usepackage{verbatim}
\usepackage{hyperref}
\usepackage{enumerate}

\begin{document}

\title{Band's Geometry Origin of Quantum Spin Transport Phenomena}

\author{Elena Derunova$^{1,2,3}$}
\author{Mazhar N. Ali$^{1,2}$}

\affiliation{$^1$ Delft University of Technology, Delft, Netherlands
$^2$ Max Plank Institute of Microstructure Physics, Halle, Germany
$^3$ Leibniz Institute for Solid State and Materials Research IFW Dresden, Germany
}

\date{\today}

\begin{abstract}
We develop a geometric description of spin-dependent transport based on the local structure of electronic bands and the Fermi surfaces. For quasi-two-dimensional systems, we show that hyperbolic regions of constant-energy surfaces generate a geometrical contribution to the Fermi velocity that couples naturally to electron spin and produces a spin-current response. We further show that, in the presence of time-reversal symmetry, the algebra of spin operators can be related to the exterior algebra of the band's tangent space, providing an additional geometric interpretation of spin in momentum space. This framework motivates a symplectic description of spin-separated transport on Fermi surfaces and its extension to three-dimensional band manifolds through contact geometry. Our results establish a direct connection between Fermi-surface geometry and intrinsic spin transport.

\end{abstract}


\maketitle

\section{Introduction}

\vspace{2mm}


The generation and control of spin currents lie at the heart of modern spintronics \cite{zutic_spintronics_2004, wolf_spintronics_2001}. A paradigmatic example is the spin Hall effect (SHE), where an applied electric field produces a transverse spin current \cite{dyakonov_she_1971, hirsch_she_1999, murakami_she_2003, sinova_she_2004}. Traditionally, intrinsic SHE is understood through the spin Berry curvature of occupied bands \cite{xiao_berry_2010, guo_intrinsic_she_2008}, requiring computationally intensive ab initio calculations and dense momentum-space integration \cite{yao_first_principles_2004, qiao_she_wannier_2018}. While successful, this approach provides limited physical intuition and little guidance for identifying new high-SHE materials efficiently \cite{sinova_she_rew, gradhand_she_review_2012}.

\vspace{2mm}


Here, we introduce a fundamentally new perspective: spin transport phenomena, including the intrinsic spin Hall response, can be directly connected to the local geometry of electronic bands and the corresponding Fermi surfaces. We show that the out-of-plane spin currents originate from hyperbolic regions of the constant-energy surfaces, like the Fermi surface. Crucially, this link allows us to predict spin currents from the Fermi surface geometry without computing the full spin Berry curvature, relying instead on a geometrical analysis of the Fermi surface. To our knowledge, no prior work has established a direct connection between Fermi surface geometry and spin Hall transport.

\vspace{2mm}


Our approach is based on separating semiclassical electron velocity \cite{sundaram_wavepacket_1999, chang_berry_1995} into contributions from the external field and from geometrical momentum fluctuations on constant-energy surfaces. In regions of nontrivial geometry (i.e. non-elliptic or non-euclidean), these fluctuations give rise to effective out-of-plane currents, which can be represented as a semiclassical projection of the spin operator onto the hyperbolic velocity components, i.e. spin currents \cite{shi_proper_spin_current_2006}.

Moreover, we establish a theorem stating that, in the presence of time-reversal symmetry \cite{kramers_1930, wigner_group_1959}, the electron spin can be represented as an element of the exterior (Grassmann) algebra \cite{nakahara_geometry_2003, frankel_geometry_2012} of the tangent space of the band. This provides a geometrical realization of spin in terms of differential forms defined in momentum space. This result constitutes a fundamentally new route to a definition of spin within a semiclassical framework for electronic transport effects in materials. Unlike conventional approaches, where spin is introduced as an additional  internal quantum degree of freedom \cite{dirac_principles_1958}, our formulation identifies spin as an intrinsic geometrical object associated with the local structure of the band.

\section{Results}


\vspace{2mm}

\vspace{2mm}

\textit{Fermi velocity for hyperbolic Fermi surfaces}. 

The mean velocity of an electron  in the periodic lattice with momentum k can be found in the semiclassical approach as \cite{ashcroft_solid_2011}:

\begin{equation}\label{vel} 
v_n(k)=\frac1 \hbar \nabla_k\varepsilon_n (k)
\end{equation}

where $\varepsilon_n (k)$ denotes the n-th eigenvalue of the energy operator. For pure 2D systems $v_n(k)=v^{xy}_n(k)=(v_x,v_y)$ is in-plane velocity. For quasi 2D systems like thin films, out-of-plane motion is still possible and thus the 3d dimension needs to be included in the model. However, in case of motion in the 3d dimension in real space, the periodicity condition is not enough to manifest the quasi-continuity of $\varepsilon_n (k)$ in the 3d dimension and apply formula \ref{vel} in the 3D space. Nevertheless, when considering the electric field to act in the xy-plane, the generated semiclassical change of momentum remains also in-plane (e.g. $dk_x/dt= (eE_x)1/\hbar$ \cite{ashcroft_solid_2011}). \textit{Then, inclusion of the 3d dimension is possible by assuming an implicit local dependence   $k_z=k_z(k_x,k_y)$ modeling quasi-2D case}. Such a parametrization can be found e.g. from the $\varepsilon_n (k)=const$ condition. Below will be shown how this lead to the out-of-plane component of the velocity. 

\vspace{2mm}



For a quasi-2D case, the velocity vector can be written as:

\begin{equation}\label{vel_split}
v_n(k)=v_{ext}+v_{geom},
\end{equation}

where $v_{ext}=v_{xy}$ is a pure 2D velocity due to the external electromagnetic potential in the $xy$-direction with corresponding non-zero $dk_{xy}$.  $v_{geom}$ is the velocity generated by the fluctuation of the momentum around constant energy surface, i.g.  
$k_z$ changes only within the surface $k_z=(k_x,k_y)$ around an infinitesimally small neighborhood of the $k_{z_0}=k_z(k_{x_0},k_{y_0})$, so that actual continuity of the constant energy surface in the z-direction is not needed. 
The origin of $v_{geom}$ is interesting: unlike the in-plane $dk_{xy}$ generated by the external electro-magnetic potential, the fluctuations of momentum on the constant energy do not need any external potential and basically are related to the quasiparticles forming on such surfaces. 

\vspace{2mm}

Anyway, both velocities are defined via \ref{vel}, but since $dk$ in \ref{vel} is generated by two different sources with non-simultaneous dynamics on different energy scales, i.e. $dk_{xy}^{ext}>>dk_{xy}^{geom}$, the velocities $v_{ext}$ and $v_{geom}$  should be found separately via the approach \ref{vel}. Basically, such a difference in energy scale validates the Fermi liquid approach of using \ref{vel} also for quasiparticles with $dk^{ext}_{xy}$ neglecting $dk^{geom}_{xy}$ in pure 2D case. However, in quasi 2D case if $k_z=(k_x,k_y)\backsim o(|k|)$ then  $dk^{geom}_{xy}<<dk_z^{geom}$. Therefore $dk^{ext}_{xy}\backsim dk_z^{geom}$, making $v^{z}_{geom}$ comparable to $v_{ext}$, brings $v_z=\frac{\partial\varepsilon_n}{\partial_{k_z}}$ implicitly into the model, and allows us by obeying the rule $\frac{\partial\varepsilon_n}{\partial_{k_z}}=\frac{\partial\varepsilon_n}{\partial_{k_z}} \nabla_{k_x k_y}\ k_z$ to express $v_{geom}$ as the following: 

\begin{equation}\label{vel_part} 
\begin{gathered}
v^{}_{geom}(k)=\frac1 \hbar (v_z\frac{\partial {k_z}}{\partial_{k_x}},v_z\frac{\partial {k_z}}{\partial_{k_y}}) P_z
\end{gathered}
\end{equation}

\vspace{2mm}
where $P_z$ is a projector to the $k_z$ directoin. 
\vspace{2mm}

Let's consider the condition $k_z=(k_x,k_y)\backsim o(|k|)$ more closely. In a normal metal $\varepsilon (k)=m_xk_x^2+m_yk_y^2+m_zk_z^2$, which would not lead to any $dk_z$. However, e.g. in  Weyl semimetals $\varepsilon_{\pm}(k)
=
\pm\sqrt{
\gamma^2(k_x^2-m)^2+v^2k_y^2+v^2k_z^2
}$ \cite{Okugawa2014} leads to $k_z\backsim |k|^2$,  which makes the velocity \ref{vel_part} relevant.  
\vspace{2mm}

Since the presence of both inversion ($\mathcal{I}$) and time-reversal ($\mathcal{T}$) symmetries is essential for the stabilization of Weyl nodes, let us consider the more general effect of these symmetries on the algebraic-geometric structure of the dispersion relation and the rise of the geometric velocity component \ref{vel_part}. The combined space-time inversion operator $\mathcal{I}\mathcal{T}$ acts on the system by imposing a global physical equivalence between the states at momentum $k$ and $-k$ across the Brillouin zone. Consequently, this quotient mapping by the $\mathbb{Z}_{2}$ symmetry group allows us to identify the effective topology of the state manifold with that of a projective space structure. 
\vspace{2mm}

Within this projective geometric framework, the complexified dispersion relation behaves as an algebraic curve whose topological genus obeys the genus-degree formula:
$g = \frac{(d-1)(d-2)}{2}
$ \cite{hartshorne1977algebraic}. Going beyond the standard Weyl model where $k_z \sim |k|^2$, let us consider a higher algebraic degree $d \ge 3$: the genus of the corresponding constant energy surface is strictly constrained to $g \ge 1$. This inherently dictates that its Euler characteristic satisfies $
\chi = 2 - 2g \le 0
$
According to the Gauss-Bonnet theorem \cite{docarmo2016differential}$\int_{M} K \, dA = 2\pi \chi(M) \le 0
$
meaning that \textit{the corresponding constant energy surface must have regions with negative Gaussian curvature $K$, i.e. hypebolic regions}.

\vspace{2mm}
For the hyperbolic regions of the constant energy surface parametrized by  $k_z=(k_x,k_y)$   $sign \frac{\partial {k_z}}{\partial_{k_x}} \neq  sign \frac{\partial {k_z}}{\partial_{k_y}}$, assume  $\frac{\partial {k_z}}{\partial_{k_x}}<0 , \frac{\partial {k_z}}{\partial_{k_y}}>0$ . Then we can rewrite (\ref{vel_part})  as the following: 

\begin{equation}\label{v_hyp}
 v^{xy}_{hyp}(k)= \frac {2} {\hbar^2}  v_{z} S_z\nabla_{k_xk_y}k_z
\end{equation}

This expression includes a spin current operator's out-of-plane component $J^i_j=\frac 1  2 (S_i,v_j)$ and thus should be relevant to the generated out-of-plane spin currents. To obtain the semiclassical spin current expression let's start with the following formula:

\begin{equation}
\mathbf{j} = -\frac{e}{(2\pi)^3}\int_{occupied} {{v}_n(k) dk}
\end{equation}

Substituting the velocity expression from equation (\ref{v_hyp}) we obtain
\begin{equation}\label{j_shc}
\mathbf{j}^{spin}_z = S_z \mathbf{j}^{hyp}_{xy},
\end{equation}

i.e. the generated by \ref{v_hyp} current is a projection of the spin operator into the "hyperbolic current", defined by: 

\begin{equation}\label{j_hyp}
\mathbf{j}^{hyp}_{xy} = -\frac{e}{(2\pi)^2\hbar}\int_{occupied} I(k) 2 v_z  \nabla_{k_xk_y}(k_z) dk 
\end{equation}


$I(k)$ is a 1 if the corresponding constant energy surface is hyperbolic and 0 otherwise. For a spinless system $S_z = 0$ , but for spin $\frac1 2$ electron $S_z \neq 0$ and thus current defined by the equation \ref{j_shc} is  non-zero manifesting as the intrinsic SHE. The SHE in this case is proportional to  $\nabla_{k_xk_y} (k_z)$, which corresponds to a geometrical curvature of a constant energy surface. Since $\nabla_{k_xk_y} (k_z)dk=ds$ integration can be reduced via the Stocks theorem to the FS integration. In our previous work we tested this approach numerically and it correlates extremely well ($R^2=0.95$) with the SHC calculeted via spin Berry curvature formalism \cite{derunova_FSdescriptor}.
\vspace{2mm}

Such a simple representation of a spin current is not an arbitrary result but rather shows a deeper fundamental mathematical connection between the geometrical objects in the k-space and electron's spin as will be shown below.

\vspace{2mm}

It is important to emphasize that the aforementioned symmetries and the specific high algebraic degree of the dispersion ($d \ge 3$) constitute a set of sufficient, but not necessary, conditions for the emergence of hyperbolic points on the constant energy surface. Local hyperbolic regions can readily manifest under less restrictive terms. While our algebraic-geometric framework provides a robust global pathway to negative Gaussian curvature, it does not preclude the existence of hyperbolic dynamics in lower-symmetry or linear-dispersion regimes with the same additional velocity component.

\vspace{2mm}
\textit{Spin algebra on the band}. 
On one hand side, geometrically $v=(\frac{\partial\varepsilon_n}{\partial_{k_x}},\frac{\partial\varepsilon_n}{\partial_{k_y}},\frac{\partial\varepsilon_n}{\partial_{k_z}})$ is the normal vector of the constant energy surface. On the other hand, a linear hull of its components generates a space known as the tangent space $T_k\varepsilon_n = span<\frac{\partial\varepsilon_n}{\partial_{k_x}},\frac{\partial\varepsilon_n}{\partial_{k_y}},\frac{\partial\varepsilon_n}{\partial_{k_z}}>$, which is itself a real 3D euclidian vector space, i.e. it's isomorphic to $\mathbb{ R}^3$. In other words, there are two 3D spaces (k-space and $T_k\varepsilon_n$) intersecting at the normal vector $v=(\frac{\partial\varepsilon_n}{\partial_{k_x}},\frac{\partial\varepsilon_n}{\partial_{k_y}},\frac{\partial\varepsilon_n}{\partial_{k_z}})$. 
With the elements of $T_k\varepsilon_n$ one can build a Grasmannian algebra $\Lambda (T_k\varepsilon_n)$, which is an algebra of multiplication of elements from $T_k\varepsilon_n$ to k-vectors \cite{nakahara2003geometry,lee2013smooth}. Below we show how this algebra relates to the electron spin.

\vspace{2mm}
\textbf{Theorem}. In the presence of time-reversal symmetry, spin can be presented as an element from $\Lambda T_k (\varepsilon_n)$.  
\vspace{2mm}

\textit{Proof.} Consider now a spinful system, where $\psi_{m, k}$ represents spin up and $\psi_{m+1, k}$ spin down. One can equip the k-space with a quadratic form. 

\begin{equation}
Q_{m,n}(k)=<\psi_{m, k};\psi_{n, k}>
\end{equation}

In the presence of time-reversal symmetry a band will be degenerate due to the Kramers theorem \cite{kramers_1930}, i.e. $\varepsilon_m(k)=\varepsilon_{m+1}(k)$ . Then the form $Q_{m,m+1}$ is zero at every point in the k-space (which is  $\mathbb{ R}^3$ as the momentum space). Then the corresponding  Clifford algebra $Cl(\mathbb{ R}^3, Q_{m,m+1})$ is isomorpohic to the algebra generated by the Pauli matrices (by construction describing the spin variable)  \cite{lounesto2001clifford,porteous1995clifford}:
$$ Alg(\sigma_x,\sigma_y,\sigma_z  ) \simeq Cl (\mathbb{ R}^3, Q_{m,m+1})  $$

Clifford algebra itself is isomorphic to the Grassman algebra of the $\mathbb{ R}^3$ k-space \cite{chevalley1997clifford,lounesto2001clifford}: 

$$  Cl (\mathbb{ R}^3, Q_{m,m+1}) \simeq \Lambda (\mathbb{ R}^3)$$

Since the tangent space of the band $T_k\varepsilon_{m}$ is isomorphic to $\mathbb{ R}^3$ (considering a spinless band, without loss of generality, to be equivalent to the m-th band) we have  the following  equivalence: 

\begin{equation}\label{spin_equiv}
Alg(\sigma_x,\sigma_y,\sigma_z)  \simeq \Lambda(T_k\varepsilon_{m})   
\end{equation} 


\vspace{2mm}

The spin projection operator is defined as $S_i=\frac\hbar 2 \sigma_i$ \cite{sakurai2017modern,griffiths2018quantum} and due to \ref{spin_equiv} $\sigma_i$ can be replaced to an object from the $\Lambda(T_kM_{m\  band})$. Thus we showed that spin of a spinless band can be constructed via external algebra of the tangent space of the band.\qed

\vspace{2mm}

Even though time reversal symmetry was assumed for such a geometrical interpretation of the spin, this approach in principle can be extended to the non time reversal symmetric case involving more advanced geometrical constructions as will be shown below.

\vspace{2mm}

\textit {Symplectic structure on the Fermi surface}. All statements in this section are of a conjectural nature and should be regarded as hypotheses. For the simplicity consider first the Fermi surface of the free electrons, e.g. the Fermi sphere \cite{ashcroft_solid_2011, ziman1972principles}. Spinful electrons would still occupy the same Fermi sphere due to the Pauli exclusion principle. The semiclassical approach, however, does not distinguish between the up and the down states in the calculation of current, but assumes both states to contribute equally to the total conductance \cite{ashcroft_solid_2011,xiao_berry_2010}. To overcome this we can introduce a natural geometric separation of two states via the sides (orientation) of the Fermi sphere which is consistent with the different directions of semiclassical dynamics along the FS in the magnetic field ($dk/dt=-e/\hbar^2(\nabla_k\varepsilon \times B)$) \cite{ashcroft_solid_2011}. This way allows also to change the view on the integration along the FS, e.g. in the caclulation of the semiclassical current via $j=\int_{FS}...ds$, changing $ds$ to an oriented area element, i.e. the symplectic form $d\omega$ \cite{mcduff2017introduction,cannas2008lectures}. It has a standard form in local FS coordinates:

\begin{equation} \label{sympl_gen}
    \omega=dk_\nu\wedge dk_\mu
\end{equation}
In the global coordinates of the reciprocal space it takes the following form:
\begin{equation} \label{sympl_coord}
    \omega=\sum_{\text{cyclic}(x,y,z)}
k_x \, dk_y \wedge dk_z
\end{equation}

Considering a particular parameterization e.g. $k_z=k_z(k_x,k_y)$, i.e., separating states only with regard to the out of $xy$ plane spin, reduces this to the form $k_z dk_x\wedge dk_y$ appearing in \ref{v_hyp}. 

Acoording to the Daurboux theorem the symplectic structere always take form \ref{sympl_gen}  \cite{mcduff2017introduction,arnold1989mathematical} and thus it can be used for any FS of real materials, which is different from sphere, to claculate spin-separated transport from the FS of spin-degenerate states. However this approach cannot be directly used for the calculation of the transport from the band, since band is a 3D manifold and symplectic structures exist only on even-dimensional manifolds \cite{arnold1989mathematical,cannas2008lectures}. The standard possibility to introduce a symplectic structure is to double dimensionality by adding tangent space of the band, as been considered in \ref{spin_equiv}. This approach cannot guarantee correctness for not time-reversal symmetric case. 

Instead, in 3D case we can consider an odd-dimensional version of symplectic geometry, called contact geometry. In this case the following standard form \cite{geiges2008introduction,arnold1989mathematical}: 

\begin{equation} \label{cont_form}
    \omega=k_xdk_y+dk_z
\end{equation}

Represents contact structure, i.e. distribution of planes in the k-space to which $\varepsilon (k)$ will be maximailly non-integrable. For the form \ref{cont_form} it is planes spinning along the $k_y$ direction. This naturaly has similar meaning to the spin of the electron, and appears as a non-integrability of the system purely due to the geometry (dimensionality) of the space.



\vspace{2mm}

 \textit{Semi-metalic Fermi surfaces}. The normals to the Fermi surface correspond not only to orientation of the FS orbits under applied magnetic field but also to the direction of the the Fermi velocity: directed outward normals represent pure electron-like sheets and directed inward represent hole-like sheets in metals\cite{ashcroft_solid_2011}  Thus for pure metalic electronic structure the Fermi surface can't have local hyperbolic points, only the global hyperbolic points resulting into the open orbits on the Fermi surface. In contrasts to this the semi-metalic compounds may have such local hyperbolic points even on the closed pockets of the Fermi surface. These points identify the band anti-crossings like Dirac points belonging to the valence band and the conductance band at the same time and realizing the "flipping sides" on the semimetalics Fermi surfaces. Thus the Fermi surface at this points should  have degenerate normals, which is possible in case if they are locally hyperbolic or if pure electron pocket is connected with pure hole-pocket by symmetry at the edges of BZ, which can be identified though the presence of globally hyperbolic points. Therefore we can apply a similar symplectic structure based approach to predict anomalous transport in semi-metals, e.g. the method \ref{j_shc} show numerical correlation with the Anomalous Hall Effect (AHE) as well \cite{derunova_FSdescriptor}.  

\subsection{DISCUSSION AND CONCLUSION}

We have shown that hyperbolic regions of the Fermi surface generate an additional geometric contribution to the Fermi velocity, and that this contribution, projected onto the spin operator, produces an intrinsic out-of-plane spin current. Because the existence of these hyperbolic regions follows from a topological argument (the genus--degree formula combined with the Gauss--Bonnet theorem) rather than from a case-by-case numerical search, this construction provides a simple, physically transparent semiclassical formula that correlates with the spin Hall conductivity without requiring a full spin Berry curvature calculation.

\vspace{2mm}

A deeper consequence of this result concerns the relation between spin and band geometry itself. We proved that, under time-reversal symmetry, the spin operator can be represented as an element of the exterior algebra of the band's tangent space, via the isomorphism between the Clifford algebra of the Pauli matrices and the Grassmann algebra of $\mathbb{R}^3$. This result shows that spin need not be introduced as an independent internal quantum number: at least in the time-reversal-symmetric case, it can be recovered from the local geometric structure of the band. In this sense, the momentum-space geometry that governs the semiclassical velocity and the algebraic structure that governs spin are two aspects of the same underlying object, the tangent space of the band.This geometric picture also suggests an interpretive analogy worth noting, though we emphasize it as a heuristic rather than a derived result. Spin may be viewed as encoding a geometric correspondence between velocity in real space and momentum in reciprocal space, two descriptions that the uncertainty principle keeps formally distinct, with spin furnishing the geometrical structure relating them as a basis of the corresponding Hilbert space.

\vspace{2mm}

The symplectic and contact-geometric constructions we introduce to extend spin-separated transport beyond the time-reversal-symmetric case are, at this stage, conjectural, and we present them as a candidate framework rather than a completed theory. In particular, the special role played by the sphere $S^2$ --- the Fermi surface of free electrons --- suggests a possible link between the existence of spin and the three-dimensionality of momentum space, because $S^2$ is the only sphere out of all n-dimensional spheres admitting the symplectic structure. However, establishing this rigorously, as a counterargument to the string theories is left for future work.

\vspace{2mm}

Taken together, these results indicate that the geometry of the Fermi surface is not merely a passive backdrop for computing transport coefficients, but an active ingredient that determines both the existence and the magnitude of intrinsic spin currents. We expect this geometric perspective to be useful both as a practical, low-cost screening tool for materials with strong spin Hall response and as a starting point for a more general geometric theory of spin-dependent transport phenomena, including possible extensions to superconducting pairing, spin textures etc.

\bibliographystyle{ieeetr}
\bibliography{lib2}

\end{document}